\documentstyle[aas2pp4,amsmath]{article}

\def\hii{H\thinspace{$\scriptstyle{\rm II}$}~} 
 
\begin{document}

\lefthead{Steigman, Hata, and Felten} 
\righthead{Non-Nucleosynthetic Constraints on the Baryon Density...} 
 
\rightline{May 28, 1998 (revised)} 
\rightline{OSU-TA-14/97} 
\rightline{IASSNS-AST 97/43} 
 
\title{          Non-Nucleosynthetic Constraints on the Baryon Density  \\ 
                 and Other Cosmological Parameters   } 
\author{         Gary Steigman\altaffilmark{1,2}, 
                 Naoya Hata\altaffilmark{3}, and 
                 James E.\ Felten\altaffilmark{4,1}       } 
 
\altaffiltext{1}{Department of Physics, The Ohio State University,  
174 West 18th Avenue, Columbus, OH 43210} 
\altaffiltext{2}{Department of Astronomy, The Ohio State University; steigman@mps.ohio-state.edu} 
\altaffiltext{3}{Institute for Advanced Study, Princeton, NJ 08540; hata@ias.edu} 
\altaffiltext{4}{Code 685, NASA Goddard Space Flight Center, Greenbelt, MD 20771; 
felten@stars.gsfc.nasa.gov (present address)}

\begin{abstract} 
 
Because the baryon-to-photon ratio $\eta_{10}$ is in some doubt, we  
drop nucleosynthetic constraints on $\eta_{10}$ and fit the three  
cosmological parameters $(h, \Omega_{\mathrm{M}}, \eta_{10})$ to four  
observational constraints: Hubble parameter $h_{\mathrm{o}} = 0.70 \pm  
0.15$, age of the universe $t_{\mathrm{o}} = 14^{+7}_{-2}$ Gyr,  
cluster gas fraction $f_{\mathrm{o}} \equiv f_{\mathrm{G}}h^{3/2} =  
0.060 \pm 0.006$, and effective shape parameter $\Gamma_{\mathrm{o}} =  
0.255 \pm 0.017$.  Errors quoted are $1\sigma$, and we assume Gaussian 
statistics.  We experiment with a fifth constraint  
$\Omega_{\mathrm{o}} = 0.2 \pm 0.1$ from clusters.  We set the tilt  
parameter $n = 1$ and the gas enhancement factor $\Upsilon = 0.9$.   
We consider CDM models (open and $\Omega_{\mathrm{M}} = 1$) and flat  
$\Lambda$CDM models.  We omit HCDM models (to which the $\Gamma_ 
{\mathrm{o}}$ constraint does not apply).  We test goodness of fit and  
draw confidence regions by the $\Delta\chi^2$ method.  CDM models with  
$\Omega_{\mathrm{M}} =1$ (SCDM models) are accepted only because the  
large error on $h_{\mathrm{o}}$ allows $h < 0.5$.  Baryonic matter  
plays a significant role in $\Gamma_{\mathrm{o}}$ when  
$\Omega_{\mathrm{M}} \sim 1$. Open CDM models are accepted only for  
$\Omega_{\mathrm{M}} \gtrsim 0.4$.  The combination of the four other  
constraints with $\Omega_{\mathrm{o}} \approx 0.2$ is rejected in CDM  
models with 98\% confidence, suggesting that light may not trace mass.  
$\Lambda$CDM models give similar results.  In all of these models,  
$\eta_{10}$ $\gtrsim 6$ is favored strongly over $\eta_{10}$ $\lesssim  
2$.  This suggests that reports of low deuterium abundances on QSO  
lines of sight may be correct, and that observational determinations  
of primordial $^4$He may have systematic errors.  Plausible variations  
on $n$ and $\Upsilon$ in our models do not change the results much.  
If we drop or change the crucial $\Gamma_{\mathrm{o}}$ constraint,  
lower values of $\Omega_{\rm M}$ and $\eta_{10}$ are permitted. The  
constraint $\Gamma_{\mathrm{o}} = 0.15 \pm 0.04$, derived recently  
from the IRAS redshift survey, favors $\Omega_{\rm M} \approx 0.3$  
and $\eta_{10} \approx 5$ but does not exclude $\eta_{10} \approx 2$.

\end{abstract} 

\section{INTRODUCTION} 
\label{Sec:Introduction} 
 
In a Friedmann-Lema\^\i tre big bang cosmology, the universal baryonic  
mass-density parameter $\Omega_{\mathrm{B}}\; (\,\equiv 8 \pi G  
\rho_{\mathrm{B}}/3H_0^2\,)$ may be calculated from  
\begin{equation}  
\begin{split}  
\Omega_{\mathrm{B}}\,h^2  
& 
= 3.675 \times 10^{-3}(T/2.73\,\mathrm{K})^3 \;  
\eta_{10} \\   
& 
= 3.667 \times 10^{-3} \; \eta_{10},  
\label{Eq:Omega_B}  
\end{split}  
\end{equation}  
where $h$ is defined by the present Hubble parameter $H_0 \; [\, h  
\equiv H_0/(100$ km s$^{-1}$ Mpc$^{-1})\,]$, $T$ is the present  
microwave background temperature, and $\eta_{10}$ is the  
baryon-to-photon number ratio in units $10^{-10}$.  The last member of  
equation~(\ref{Eq:Omega_B}) is obtained by setting $T = 2.728$ K  
(Fixsen et al. 1996).  In principle, $\eta_{10}$ is well determined  
(in fact overdetermined) by the observed or inferred primordial  
abundances of the four light nuclides D, $^3$He, $^4$He, and $^7$Li,  
if the number of light-neutrino species has its standard value $N_\nu  
=3$.  For some years it has been argued that $\eta_{10}$ is known to  
be $3.4 \pm 0.3$ (Walker et al. 1991; these error bars are about  
``1$\sigma$''; cf.  Smith, Kawano, \& Malaney 1993) or at worst $4.3  
\pm 0.9$ (Copi, Schramm, \& Turner 1995a; cf. Yang et al. 1984), and  
that equation~(\ref{Eq:Omega_B}) is a powerful constraint on the  
cosmological parameters $\Omega_{\mathrm{B}}$ and $h$.  
  
In practice, it seems recently that $\eta_{10}$ may not be so well  
determined, and even that the standard theory of big bang  
nucleosynthesis (BBN) may not give a good fit.  With improved  
abundance data, it appears that the joint fit of the theory to the  
four nuclide abundances is no longer good for any choice of  
$\eta_{10}$ (Hata et al. 1995).  These authors offer several options  
for resolving the apparent conflict between theory and observation.  
Although they suggest that some change in standard physics may be  
required (e.g., a reduction in the effective value of $N_\nu$ during  
BBN below its standard value 3), they note that large systematic  
errors may compromise the abundance data (cf. Copi, Schramm,  
\& Turner 1995b).  The nature of such errors is unclear, and this remains   
controversial.  Other authors have reacted to the impending crisis in  
self-consistency by simply omitting one or more of the four nuclides  
in making the fit (Dar 1995; Olive \& Thomas 1997; Hata et al. 1996,  
1997; Fields et al. 1996).  
  
This controversy has been sharpened by new observations giving the  
deuterium abundances on various lines of sight to high-redshift QSOs.  
These data should yield the primordial D abundance, but current  
results span an order of magnitude.  The low values, D/H by number  
$\approx 2 \times 10^{-5}$ (Tytler, Fan, \& Burles 1996; Burles \&  
Tytler 1996), corresponding to $\eta_{10} \approx 7$ in the standard  
model, have been revised slightly upward [D/H $\approx (3-4) \times  
10^{-5}$ (Burles \& Tytler 1997a,b,c); $\eta_{10} \approx 5$], but it  
still seems impossible to reconcile the inferred abundance of $^4$He  
[Y$_{\rm P} \approx 0.234$; Olive \& Steigman 1995 (OS)] with standard  
BBN for this large value of $\eta_{10}$ (which implies Y$_{\rm BBN}  
\approx 0.247$) unless there are large systematic errors in the $^4$He  
data (cf. Izotov, Thuan, \& Lipovetsky 1994, 1997).  Such low D/H values  
have also been challenged on observational grounds by Wampler (1996) and  
by Songaila, Wampler, and Cowie (1997), and deuterium abundances nearly  
an order of magnitude higher, D/H $\approx 2\times10^{-4}$, have been  
claimed by Carswell et al. (1994), Songaila et al. (1994), and Rugers  
and Hogan (1996) for other high-redshift systems with metal abundances  
equally close to primordial.  Although some of these claims of high  
deuterium have been called into question (Tytler, Burles, \& Kirkman  
1997), Hogan (1997) and Songaila (1997) argue that the spectra of other  
absorbing systems require high D/H (e.g., Webb et al. 1997).  If these  
higher abundances are correct, then D and $^4$He are consistent with  
$\eta_{10} \approx 2$, but modellers of Galactic chemical evolution  
have a major puzzle: How has the Galaxy reduced D from its high primordial  
value to its present (local) low value without producing too much $^3$He  
(Steigman \& Tosi 1995), without using up too much interstellar gas  
(Edmunds 1994, Prantzos 1996), and without overproducing heavy elements  
(cf. Tosi 1996 and references therein)?  It appears that $\eta_{10}$,  
though known to order of magnitude, may be among the less well-known  
cosmological parameters at present.  Despite this, large modern simulations  
which explore other cosmological parameters are often limited to a single  
value of $\eta_{10} = 3.4$ (e.g., Borgani et al. 1997).  
  
In this situation it may be instructive, as a thought experiment, to  
abandon nucleosynthetic constraints on $\eta_{10}$ entirely and ask:  
If we put $\eta_{10}$ onto the same footing as the other cosmological  
free parameters, and apply joint constraints on all these parameters  
based on other astronomical observations and on theory and simulation,  
what values of $\eta_{10}$ and the other parameters are favored?  This  
may indicate the most promising avenue to a resolution of the  
controversy over $\eta_{10}$.  
  
We discuss the following popular CDM models: (1) Open or closed cold  
dark-matter model with cosmological constant $\Lambda = 0$ (CDM  
model).  The ``standard'' (flat) CDM model (SCDM), which is an  
Einstein-de Sitter model, is covered as a special case of this.  (2)  
Flat CDM model with nonzero $\Lambda$ ($\Lambda$CDM model).  In a flat  
model with both hot and cold dark matter, with $\Lambda = 0$ (HCDM  
model), the constraints will be different; we defer these HCDM models  
to a later paper.  Nonflat models with nonzero $\Lambda$ are not necessarily  
ruled out by ``fine-tuning" arguments and may be of interest (Steigman \&  
Felten 1995), but at the moment we are not compelled to resort to these.

Our approach will be to let three parameters range freely, fit the  
constraints (observables) other than nucleosynthetic constraints,  
test goodness of fit by $\chi^2$, and draw formal confidence regions  
for the parameters by the usual $\Delta\chi^2$ method.  Because statistical  
results of this kind are sometimes controversial, we intend to keep the work  
conceptually simple, review the constraints in a helpful way, and discuss  
our method carefully.  Error bars are $\pm 1\sigma$ unless stated otherwise.   
The $\Delta\chi^2$ approach is revealing because, in the linear approximation, 
the confidence regions obtained are rigorous as probability statements and 
require no ``a priori" probability assumptions about the unknown parameters. 
Most of our results are not surprising, and related work has been done before  
(Ostriker \& Steinhardt 1995, White et al. 1996, Lineweaver et al. 1997, White  
\& Silk 1996, Bludman 1997), but not with these three free variables and the  
full $\chi^2$ formalism.  It  
is well known that recent cosmological observations and simulations,  
particularly related to the ``shape parameter'' $\Gamma$ and the  
cluster baryon fraction (CBF), pose a challenge to popular models, and  
that there is some doubt whether any simple model presently fits all  
data well.  Our work, which begins by discarding nucleosynthetic  
constraints, provides a new way of looking at these problems.  The CBF  
and $\Gamma$ constraints have not been applied jointly in earlier work.  
We find that, given our conservative (generous) choice of error bar on  
$h$, the SCDM model is disfavored somewhat but by no means excluded,  
if we are willing to accept $\eta_{10} \gtrsim 9$. But even with the 
generous error on h, and allowing $\Omega_{\rm M}$ to range freely, large  
values ($\gtrsim 5$) of $\eta_{10}$ are favored over small values 
($\lesssim 2$). This suggests that the low D abundances measured by 
Burles and Tytler may be correct, and that the observed (extrapolated)  
primordial helium-4 mass fraction [$Y_{\mathrm{P}} \approx 0.23$; cf.  
OS and Olive, Skillman, \& Steigman 1997 (OSS)], thought to be well  
determined, may be systematically too low for unknown reasons.  
  
\section{CDM MODELS:  PARAMETERS AND OBSERVABLES}  
\label{Sec:CDM-Models}  
  
\subsection{Parameters}

We will take the CDM models to be defined by three free parameters:  
Hubble parameter $h$; mass-density parameter $\Omega_{\mathrm{M}} = 8  
\pi\, G\, \rho_{\mathrm{M}}/3H_0^2$; and baryon-to-photon ratio $\eta_{10}$,  
related to $\Omega_{\mathrm{B}}$ by equation~(\ref{Eq:Omega_B}).  Here  
$\Omega_{\mathrm{M}}$ by definition includes all ``dynamical mass'':  
mass which acts dynamically like ordinary matter in the universal  
expansion.  It is not limited to clustered mass only.  Other free  
parameters having to do with structure formation, such as the tilt  
parameter $n$, could be added (White et al. 1996; Kolatt \& Dekel 1997;  
White \& Silk 1996), but we will try in general to avoid introducing many  
free parameters, so as to avoid generating confidence regions in more than  
three dimensions.  We will, however, show results for two values of $n$  
(1 and 0.8), and for a few alternative choices of other parameters affecting  
some of the observables.

\subsection{Observables}  
  
We will consider five observables (constraints) which have measured  
values and errors which we assume to be normal (Gaussian).  The five  
observables are: (1) measured Hubble parameter $h_{\mathrm{o}}$; (2)  
age of the universe $t_{\mathrm{o}}$; (3) dynamical mass-density  
parameter $\Omega_{\mathrm{o}}$ from cluster measurements or from  
large-scale flows; (4) gas-mass fraction $f_{\mathrm{o}} \equiv  
f_{\mathrm{G}} h^{3/2}$ in rich clusters; and (5) ``shape parameter''  
$\Gamma_{\mathrm{o}}$ from structure studies.  In much of our work  
we will drop one or another of these constraints.  
  
An observable $w_{\mathrm{o}}$ has the central value $\langle w  
\rangle$ and the standard deviation $\sigma_w$.  The theoretical  
expression for this observable is given by a known function, $w$, of  
the three free parameters.  The $\chi^2$ contribution of this  
observable is written as $\chi^2 = (\langle w  
\rangle - w)^2 / \sigma_w^2$.  This sets up the usual conditions for  
the total $\chi^2$ (which, assuming the errors are uncorrelated, is a  
sum of $\chi^2$ contributions from different observables) to find the  
confidence regions for the free parameters (Cramer 1946, Bevington \&  
Robinson 1992, Press et al. 1992, Barnett et al. 1996).  We state  
below the theoretical expression $w$ and the observational constraint  
$w_{\mathrm{o}} = \langle w \rangle \pm \sigma_w$ which we assume.  
  
There are other constraints which could be applied, including cluster  
abundance, the height of the ``acoustic peak'' in the angular  
fluctuation spectrum of the cosmic background radiation, the  
Sunyaev-Zeldovich effect in clusters, the Lyman-alpha forest, and  
theoretical constraints on $\Lambda$ (White et al. 1996; White \& Silk  
1996; Lineweaver et al. 1997; Myers et al. 1997; Rauch, Haehnelt, \& 
Steinmetz 1997; Weinberg et al. 1997; Bi \& Davidsen 1997; Fan, Bahcall,  
\& Cen 1997; Martel, Shapiro, \& Weinberg 1998).  We omit these here but  
intend to pursue them in subsequent work. 
  
\subsection{Observed Hubble Parameter $h_{\mathrm{o}}$}  
  
For the Hubble parameter the observable $h_{\mathrm{o}}$ is simply fit 
with the parameter $h$.  Measurements of $h$ still show scatter which 
is large compared with their formal error estimates (Bureau et 
al. 1996, Tonry et al. 1997, Kundi\'{c} et al. 1997, Tammann \& 
Federspiel 1997).  This indicates systematic errors.  We do not 
presume to review this subject.  To be conservative (permissive), we 
take $h_{\mathrm{o}} = 0.70 \pm 0.15$.  Some may think that a smaller 
error could be justified.  In \S 3.2 we will experiment with shrinking 
the error bar.

\subsection{Observed Age of the Universe $t_{\mathrm{o}}$}  
  
Pre-Hipparcos observations gave the ages of the oldest globular clusters  
as $t_{\mathrm{GC}} \approx 14 \pm 2$ Gyr (Bolte \& Hogan 1995; Jimenez  
1997; D'Antona, Caloi, \& Mazzitelli 1997; cf. Cowan et al. 1997, Nittler  
\& Cowsik 1997). Some of the analyses incorporating the new Hipparcos data  
derive younger ages, $t_{\mathrm{GC}} \approx 12$ Gyr (Chaboyer et al. 
1998, Reid 1997, Gratton et al. 1997). However, as Pont et al. (1998) 
and Pinsonneault (1998) emphasize, there are systematic uncertainties 
in the main-sequence fitting technique at the 2 Gyr level, and Pont et 
al. (1998) use Hipparcos data to derive an age of 14 Gyr for M92. 
          
The theoretical age for the $\Lambda = 0$ models is given by: $t = 
9.78\:h^{-1}f(\Omega_{\mathrm{M}};\Lambda$=0) Gyr [Weinberg 1972, 
equations (15.3.11) \& (15.3.20)]. The ``observed" age of the 
universe, $t_{\mathrm{o}}$, should exceed $t_{\mathrm{GC}}$ by some 
amount $\Delta t$.  The best guess for $\Delta t$ might be 0.5 -- 1 
Gyr.  Most theorists believe that $\Delta t$ must be quite small (2 
Gyr at most), but we know of no conclusive argument to prove this, and 
we do not want long-lived models to suffer an excessive $\chi^2$ 
penalty. We could treat $\Delta t$ as another free parameter, but to 
avoid this and keep things simple, we introduce asymmetric error 
bars. We believe that $t_{\mathrm{o}} = 14^{+7}_{-2}$ Gyr is a fair 
representation of $t_{\mathrm{o}}$ derived from present data on 
$t_{\mathrm{GC}}$. This allows enough extra parameter space at large 
ages to accommodate a conservative range of $\Delta t$; extremely 
large ages will be eliminated by the $h_{\mathrm{o}}$ constraint in 
any case.  The $\chi^2$ analysis will still be valid with the unequal 
error bars if we assume the $\chi^2$ contribution as $(14 - 
t)^2/\sigma_t^2$, where $\sigma_t$ takes the value 2 Gyr when $t < 14$ 
Gyr and the value 7 Gyr when $t > 14$ Gyr.

\subsection{Observed Mass Density $\Omega_{\mathrm{o}}$}

The observed $\Omega$ at zero redshift, determined from clusters, has  
recently been reported as  
\begin{equation}  
\Omega_{\mathrm{CL}} = 0.19 \pm 0.06 {\mbox { (stat)} }   
\pm 0.04 {\mbox { (sys)} }  
\label{Eq:Omega_CNOC}  
\end{equation}  
(Carlberg, Yee, \& Ellingson 1997), where the respective errors are  
statistical and systematic.  This is based on the $M/L$ ratio in  
clusters and the luminosity density of the universe.  If light traces  
dynamical mass, then we expect that $\Omega_{\mathrm{CL}}$ can be  
directly fit with $\Omega_{\mathrm{M}}$.  
  
There is a difficulty in using equation~(\ref{Eq:Omega_CNOC}) as a  
constraint on the underlying parameter $\Omega_{\mathrm{M}}$.  Many  
consumers of equation~(\ref{Eq:Omega_CNOC}) and earlier results  
(cf. Carlberg et al. 1996) have failed to notice that the result is  
model-dependent, because the clusters in the sample have substantial  
redshifts (0.17 $< z <$ 0.55).  In their analysis Carlberg et  
al. (1997) assumed $q_0 = 0.1$ (e.g., $\Omega_{\mathrm{M}} = 0.2$ and  
$\Lambda = 0$).  When the result is $\langle \Omega_{\mathrm{CL}}  
\rangle= 0.19$, clearly the analysis is approximately self-consistent,  
but this does not give us sufficient guidance in exploring other  
values of $\Omega_{\mathrm{M}}$.  For example, effects of nonzero  
$\Lambda$ could be substantial.  We believe that if the analysis of  
Carlberg et al. (1997) were repeated for a $\Lambda$CDM model, the  
resulting $\langle \Omega_{\mathrm{CL}} \rangle$ might be smaller,  
perhaps 0.12 rather than 0.19.  The parameters $\Omega_{\mathrm{M}}$  
and $\Lambda$ need to be incorporated more fully into the analysis.  
  
Looking at these and earlier data, we choose, somewhat arbitrarily, to  
use instead of equation~(\ref{Eq:Omega_CNOC}) a somewhat more permissive  
$\Omega$ constraint from clusters:  
\begin{equation}  
\Omega_{\mathrm{o}} = 0.2 \pm 0.1  
\label{Eq:Omega_CL}  
\end{equation}  
(Carlberg 1997).  If the critical (``closure'') $M/L$ ratio in  
$B_{\mathrm{T}}$ magnitude is $1500\,h\, (M/L)_{\odot}$ (Efstathiou,  
Ellis, \& Peterson 1988), then equation~(\ref{Eq:Omega_CL}) requires  
that the mean $M/L$ in $B_{\mathrm{T}}$ for galaxies in the local  
universe be about $(300 \pm 150)\: h$ (cf. Smail et al. 1996).  This  
agrees well with modern reviews (Bahcall, Lubin,  
\& Dorman 1995, Trimble 1987).  
  
We will assume in some of our examples that $\Omega_{\mathrm{o}}$ can  
be directly fit with $\Omega_{\mathrm{M}}$ (light traces mass;  
``unbiased''), with $\Omega_{\mathrm{o}}$ given by  
equation~(\ref{Eq:Omega_CL}).  Obviously, under this assumption, the  
$\chi^2$ contribution from the observed $\Omega_{\mathrm{o}}$ will  
rule out the SCDM model ($\Omega_{\mathrm{M}} = 1$) with high  
confidence.  
  
Bias is possible, with the most likely bias being $\Omega_{\mathrm{o}}  
< \Omega_{\mathrm{M}}$.  This would be the case of additional  
unclustered or weakly clustered dynamical mass.  Because such weakly  
clustered mass is quite possible, we will also do other cases with an  
alternative to the cluster constraint, as follows.  Dekel and Rees  
(1994; cf.  Dekel 1997) studied large-scale flows around voids and  
concluded that $\Omega_{\mathrm{M}}$ must be quite large:  
$\Omega_{\mathrm{M}} > (0.4, 0.3, 0.2)$ at confidence levels  
$(1.6\sigma, 2.4\sigma, 2.9\sigma)$.  All values $\Omega_{\mathrm{M}}  
\ge 0.6$ were permitted.  To use this one-way constraint as the third  
observable in a $\chi^2$ fit, we need a substitute function $\chi^2  
(\Omega_{\mathrm{M}})$ having the properties: $\chi^2(0.4) \approx  
(1.6)^2$, $\chi^2 (0.3) \approx (2.4)^2$, $\chi^2 (0.2) \approx  
(2.9)^2$.  The function  
%
%
%
%
\begin{equation}  
\chi^2 (\Omega_{\mathrm{M}}) =   
\left\{  
\begin{array}{ll}  
   (0.6 - \Omega_{\mathrm{M}})^2/(0.125)^2  &   
                            \mbox{$(\Omega_{\mathrm{M}} < 0.6)$}, \\  
   0 &                      \mbox{$(\Omega_{\mathrm{M}} \ge 0.6)$}  
\end{array}  
\right.  
\label{Eq:Omega_DR}  
\end{equation}  
is a good approximation.  For $\Omega_{\mathrm{M}} \ge 0.6$, this  
$\chi^2$ implies a ``perfect fit'' to the $\Omega$ observable.  This  
leaves parameter space open to large $\Omega_{\mathrm{M}}$.  We apply  
the Dekel-Rees constraint instead of the cluster constraint if we just  
substitute equation~(\ref{Eq:Omega_DR}) for the usual $\chi^2$ term  
arising from $\Omega_{\mathrm{o}}$.

\subsection{Observed Cluster Gas Mass Fraction $f_{\mathrm{o}}$}

Theorists and observers (White \& Frenk 1991, Fabian 1991, Briel,  
Henry, \& B\"ohringer 1992, Mushotzky 1993) have long argued that the  
large observed gas mass fraction in clusters, $f_{\mathrm{G}}$, is a  
valuable cosmological datum and poses a serious threat to the SCDM  
model.  This argument was raised to high visibility by the  
quantitative work of White et al. (1993), and now the problem is  
sometimes called the ``baryon catastrophe'' (Carr 1993) or ``baryon  
crisis'' (Steigman \& Felten 1995).  
  
At the risk of boring the experts, we must emphasize that the  
following argument does not assume that most of the mass in the  
universe, or any specific fraction of it, is in rich clusters.  
Rather, we will use $f_{\mathrm{G}}$, not as a constraint on  
$\Omega_{\mathrm{M}}$, but as a constraint on the universal baryon  
fraction, the ratio $\Omega_{\mathrm{B}}/\Omega_{\mathrm{M}}$.  The  
idea is that the content of a rich cluster is a fairly unbiased sample  
of baryonic and dark matter.  This is suggested by simulations, which  
are discussed below.  
  
The measurement of $f_{\mathrm{G}}$ poses problems, but this is not  
the place for a lengthy review.  Magnetic pressure (Loeb \& Mao 1994)  
may cause systematic errors, but these are probably not large and do  
not provide a promising escape hatch for the SCDM model (Felten 1996).  
Reported values of $f_{\mathrm{G}}$ derived by various methods show  
quite a wide range from cluster to cluster and also from groups  
through poor and rich clusters (Steigman \& Felten 1995; White \&  
Fabian 1995; Lubin et al. 1996; Mohr, Geller, \& Wegner 1996).  
Loewenstein and Mushotzky (1996) emphasize that the range in  
$f_{\mathrm{G}}$ is wider than expected from simulations, and they  
suggest that some significant physics may be missing from the  
simulations.  Cen (1997) argues that the spread may be caused by  
projection effects in the measurements of $f_{\mathrm{G}}$, arising  
because of large-scale pancakes and filaments.  Evrard, Metzler, and  
Navarro (1996), using gas-dynamical simulations to model observations,  
find that the largest error in $f_{\mathrm{G}}$ arises from  
measurement of the cluster's {\it total} mass, and that this error can  
be reduced by using an improved estimator and by restricting the  
measurement to regions of fairly high overdensity.  Evrard (1997)  
applies these methods to data for real clusters and finds  
$f_{\mathrm{G}}\,h^{3/2} = 0.060 \pm 0.003$.  This subject is still  
controversial so, to be conservative, we will double his error bars  
and take as our constraint  
\begin{equation}  
 f_{\mathrm{o}} \equiv f_{\mathrm{G}}\,h^{3/2} = 0.060 \pm 0.006.  
\end{equation}  
Note that this is quite a large gas fraction ($f_{\mathrm{G}} \approx  
17\% $ for $h \approx 0.5$), in general agreement with earlier  
results.  
  
The functional dependence of $f_{\mathrm{G}}$ on the cosmological  
parameters also poses problems.  The {\it universal} baryonic mass  
fraction is $\Omega_{\mathrm{B}}/\Omega_{\mathrm{M}}$, but not all  
baryons are in the form of gas, and furthermore selection factors may  
operate in bringing baryons and dark matter into clusters.  White et  
al. (1993) introduced a ``baryon enhancement factor'' $\Upsilon$ to  
describe these effects as they operate in simulations.  $\Upsilon$ may  
be defined by  
\begin{equation}  
f_{G0} = \Upsilon \, \Omega_{\mathrm{G}}/\Omega_{\mathrm{M}},  
\label{Eq:f_G0}  
\end{equation}  
where $\Omega_{\mathrm{G}}$ is the initial contribution of {\it gas}  
to $\Omega_{\mathrm{M}}$ (note that $\Omega_{\mathrm{G}} \le  
\Omega_{\mathrm{B}}$) and $f_{G0}$ is the gas mass fraction in the  
cluster immediately after formation.  We will shortly set $\Upsilon$  
equal to some constant.  $\Upsilon$ is really the {\it gas}  
enhancement factor, because the simulations do not distinguish between  
baryonic condensed objects if any (galaxies, stars, machos) and  
non-baryonic dark-matter particles.  All of these are lumped together  
in the term $(\Omega_{\mathrm{M}} -  
\Omega_{\mathrm{G}})$ and interact only by gravitation.  
  
If all the baryons start out as gas $(\Omega_{\mathrm{G}} =  
\Omega_{\mathrm{B}})$, and if gas turns into condensed objects only  
{\it after} cluster formation, then equation~(\ref{Eq:f_G0}) may be rewritten:  
\begin{equation}  
f_{\mathrm{G}} + f_{\mathrm{GAL}} =   
\Upsilon \,\Omega_{\mathrm{B}} / \Omega_{\mathrm{M}},  
\label{Eq:f_G+f_GAL}  
\end{equation}  
where $f_{\mathrm{G}}$ is the present cluster gas-mass fraction and  
$f_{\mathrm{GAL}}$ the present cluster mass fraction in baryonic  
condensed objects of all kinds (galaxies, stars, machos).  
We wish to carry along an estimate of $f_{\mathrm{GAL}}$ to show its  
effects.  White et al. (1993) took some pains to estimate the ratio  
$f_{\mathrm{G}}/f_{\mathrm{GAL}}$ within the Abell radius of the Coma  
cluster, counting only galaxies (no stars or machos) in  
$f_{\mathrm{GAL}}$.  They obtained  
\begin{equation}  
f_{\mathrm{G}}/f_{\mathrm{GAL}} = 5.5 \, h^{-3/2}.  
\label{Eq:f_G/f_GAL}  
\end{equation}  
This is large, so unless systematic errors in this estimate are very  
large, the baryonic content of this cluster (at least) is dominated by  
the hot gas.  Carrying $f_{\mathrm{GAL}}$ along as an indication of  
the size of the mean correction for all clusters, and solving  
equations~(\ref{Eq:f_G+f_GAL}) and (\ref{Eq:f_G/f_GAL}) for  
$f_{\mathrm{G}} h^{3/2}$, we find  
\begin{equation}  
f_{\mathrm{G}}  h^{3/2}  = [\,1 + (h^{3/2}/5.5)\,]^{-1} (\Upsilon \,  
\Omega_{\mathrm{B}}/\Omega_{\mathrm{M}})\,h^{3/2},  
\label{Eq:f_Gh}  
\end{equation}  
where $\Omega_{\mathrm{B}}$ is given from $\eta_{10}$ and $h$ by  
equation~(\ref{Eq:Omega_B}).  
This is the appropriate theoretical function of the free parameters  
to fit to the observation.    
  
The second term in brackets in equation (\ref{Eq:f_Gh}) is the small  
correction term due to $f_{\mathrm{GAL}}$.  In deriving this  
$f_{\mathrm{GAL}}$, given by equation (\ref{Eq:f_G/f_GAL}), White et  
al.\ (1993), using observations in the inner parts of bright ellipticals  
by van der Marel (1991), assumed that within cluster galaxies the mean  
ratio of baryonic mass to blue light is $6.4h(M/L)_{\odot}$. We note that  
this correction term would be larger if cluster galaxies or the cluster  
as a whole contained baryonic machos amounting to $\sim 20(M/L)_{\odot}$,  
as suggested for the halo of our Galaxy by theories of observed  
microlensing events (Chabrier, Segretain, \& M\'era 1996; Fields,  
Mathews, \& Schramm 1997; Natarajan et al. 1997).  Indeed, Gould  
(1995) has even suggested that the mass in machos could be comparable  
to that in the gas component, in which case $f_{\mathrm{GAL}} \approx  
f_{\mathrm{G}}$.

What value of the gas enhancement factor $\Upsilon$ should be used in  
equation~(\ref{Eq:f_Gh})?  A value $\Upsilon = 3-5$ would do away with  
the ``baryon catastrophe'' for the SCDM model.  There is no plausible  
way to obtain an $\Upsilon$ this large.  White et al. (1993), when  
they assumed zero-pressure gas to explore maximizing $\Upsilon$,  
always found $\Upsilon \le 1.5$ in simulations.  More realistic  
simulations with gas pressure give $\Upsilon \approx 0.9$ (Evrard  
1997), or even as small as 2/3 (Cen \& Ostriker 1993).  The gas  
preferentially stays out of the clusters to some extent rather than  
concentrating itself there.  Gas can support itself through pressure  
and shocks, while CDM cannot.  We will set $\Upsilon = 0.9$ in most of  
our examples.  This is representative of results from simulations, and  
it is close to unity, so these cases will also illustrate the  
approximate consequences if gas is neither enhanced nor excluded in  
clusters.  
  
Cen (1997) finds in simulations that the determination of  
$f_{\mathrm{G}}$ from X-ray observations is biased toward high  
$f_{\mathrm{G}}$ by large-scale projection effects; i.e., the  
calculated $f_{\mathrm{G}}$ exceeds the true $f_{\mathrm{G}}$ present  
in a cluster.  This bias factor can be as large as 1.4.  Evrard et  
al. (1996) and Evrard (1997) have not observed such a bias in their  
simulations.  If Cen is correct, we could explore the effect of such a  
bias in our statistical tests by using for $\Upsilon$, instead of 0.9,  
an ``effective value'' $\Upsilon \approx 0.9 \times 1.4 \approx 1.3$.  
Since this would also demonstrate the impact of any effect which may  
cause $\Upsilon$ to exceed unity moderately, we will show results for  
$\Upsilon = 1.3$ as well as for $\Upsilon = 0.9$.  
  
Equations~(\ref{Eq:f_G+f_GAL}) -- (\ref{Eq:f_Gh}) above were derived  
under the assumption that all baryonic condensed objects (galaxies,  
stars, machos) form from the gas {\it after} the collapse of clusters  
occurs.  If, instead, all such objects were formed {\it before}  
collapse, equation~(\ref{Eq:f_G+f_GAL}) should be replaced by:  
\begin{equation}  
f_{\mathrm{G}} + \left (   
\frac{\Upsilon - f_{\mathrm{G}}}{1 - f_{\mathrm{G}}} \right )   
f_{\mathrm{GAL}} = \Upsilon \, \Omega_{\mathrm{B}} / \Omega_{\mathrm{M}},  
\label{Eq:f_Gsum}  
\end{equation}  
reflecting the fact that the baryons in condensed objects now escape  
the gas enhancement occurring during cluster formation.  Since these  
effects are not large for $\Upsilon \approx 1$ and  
equation~(\ref{Eq:f_G+f_GAL}) is likely to be closer to the true  
situation than equation~(\ref{Eq:f_Gsum}), we will make no further use  
of equation~(\ref{Eq:f_Gsum}).

\subsection{Shape Parameter $\Gamma_{\mathrm{o}}$ from Large-Scale Structure}

The last observable we use is the ``shape parameter'' $\Gamma$, which  
describes the transfer function relating the initial perturbation  
spectrum $P_{\mathrm{I}} (k) \propto k^n$ to the present spectrum  
$P(k)$ of large-scale power fluctuations, as observed, e.g., in the  
galaxy correlation function.  When the spectral index $n$ of  
$P_{\mathrm{I}} (k)$ has been chosen, $\Gamma$ is determined by  
fitting the observed $P(k)$.  There are some notational differences  
among papers on this subject.  Sometimes $\Gamma$ is used to mean  
simply the combination $\Omega_{\mathrm{M}}h$.  We will avoid this  
usage here.  
  
Results of observations may be cast in terms of an ``effective shape  
parameter'' $\Gamma$ (White et al. 1996), which we will take as our  
observable.  Studies show that for the usual range of CDM models, with  
or without $\Lambda$, the expression for $\Gamma$ is   
\begin{equation}  
\begin{split}  
  \Gamma  
  & 
  \approx \Omega_{\mathrm{M}} h \; \exp \left [ -  
  \Omega_{\mathrm{B}} - (h/ 0.5)^{1/2} (\Omega_{\mathrm{B}} /  
  \Omega_{\mathrm{M}} ) \right ]  
  \\   
  &  
  \qquad - 0.32 \; (n^{-1} -1)  
\label{Eq:Gamma_th}  
\end{split}  
\end{equation}  
(Peacock \& Dodds 1994; Sugiyama 1995; Liddle et al. 1996a,b; White et  
al. 1996; Liddle \& Viana 1996; Peacock 1997).  For $n \approx 1$, if  
$\Omega_{\mathrm{B}}$ and $\Omega_{\mathrm{B}} /\Omega_{\mathrm{M}}$  
are small, we have $\Gamma \approx \Omega_{\mathrm{M}} h$.  The  
Harrison-Zeldovich (scale-invariant, untilted) case is $n = 1$.  We  
will adopt $n = 1$ for our standard case and experiment with different  
$n$ in \S 3.3.  Approximation (11) has been tested (and, we believe,  
is valid) only for models in which both $n$ and the exponential term  
are fairly near unity.  
	 
This $\Gamma$ is the parameter of the familiar BBKS approximation to 
the transfer function (Bardeen et al. 1986, Peacock 1997).  The BBKS 
transfer function does not fit the data continuously from long to 
short wavelengths, and Hu and Sugiyama (1996) have developed a more 
elaborate approximation for short wavelengths ($\lesssim 3h^{-1}$ 
Mpc), useful especially in cases with large $\Omega_{\rm 
B}/\Omega_{\rm M}$. Recent comparisons with data (Webster et al. 1998) 
still use Sugiyama's (1995) form for $\Gamma$ as in equation (11) 
above, and this BBKS approximation is adequate in the regime 
$(3-100)h^{-1}$ Mpc, where $\Gamma$ is determined.

For the observed value of $\Gamma$, we take  
\begin{equation}  
\Gamma_{\mathrm{o}} = 0.255 \pm 0.017  
\label{Eq:Gamma_obs}  
\end{equation}  
(Peacock \& Dodds 1994; cf. Maddox, Efstathiou, \& Sutherland 1996).  
This is based on the galaxy correlation function, and it assumes that  
light traces mass.  The very small errors, from Peacock \& Dodds  
(1994), result from averaging several data sets and may not be  
realistic.  Later we will explore the consequences of inflating these  
errors and/or moving the central value.  Equations~(\ref{Eq:Gamma_th})  
and (\ref{Eq:Gamma_obs}) imply, very roughly, that $\Omega_{\mathrm{M}}  
h \approx 0.25$.  
  
The shape parameter can be derived from the galaxy peculiar-velocity  
field instead of the density field.  The result from that technique,  
analogous to $\Omega_{\mathrm{M}} h = 0.25$, is, very roughly and for $n  
= 1$,  
\begin{equation}  
\Omega_{\mathrm{M}} h^{1.2} = 0.350 \pm 0.087 ~~(90\%   
\mbox{ CL})  
\label{Eq:Omega_Mh_Z}  
\end{equation}  
(Zaroubi et al. 1997a), where $\Omega_{\mathrm{B}} h^2 = 0.024$ has  
been assumed and CL stands for confidence level.  
Equation~(\ref{Eq:Omega_Mh_Z}) ostensibly includes an estimate of  
cosmic variance (cf. Kolatt \& Dekel 1997, Zaroubi et al. 1997b).  
Equation~(\ref{Eq:Omega_Mh_Z}) may be used to yield an estimate of  
$\Gamma_{\mathrm{o}}$ as follows: Adjust the error bar in  
equation~(\ref{Eq:Omega_Mh_Z}) from $1.65\sigma$ to $1\sigma$.  
Evaluate $\Omega_{\mathrm{M}}$ from equation~(\ref{Eq:Omega_Mh_Z}) at  
the midpoint of the ``interesting'' range of $h$, viz. $h = 0.7$.  
Substitute the resulting parameters, including $\Omega_{\mathrm{B}}$,  
into the right-hand side of equation~(\ref{Eq:Gamma_th}) and evaluate.  
The result is  
\begin{equation}  
\Gamma_{\mathrm{o}} = 0.32 \pm 0.05.  
\label{Eq:Gamma_Z}  
\end{equation}  
The independent estimates in equations~(\ref{Eq:Gamma_obs}) and  
(\ref{Eq:Gamma_Z}) agree tolerably within the stated errors.  Any  
difference, if real, could be caused by galaxy bias.  
  
The shape-parameter constraint is in a sense the least robust of the  
constraints we have discussed since it is not part of the basic  
Friedmann model.  Rather, it depends on a theory for the primordial  
fluctuations and how they evolve.  If the Friedmann cosmology were  
threatened by this constraint, we believe that those who model  
large-scale structure would find a way to discard it.  Therefore we  
will also explore some consequences of removing this constraint.

\begin{figure}[t] 
 
\epsscale{1.0} 
 
\plotone{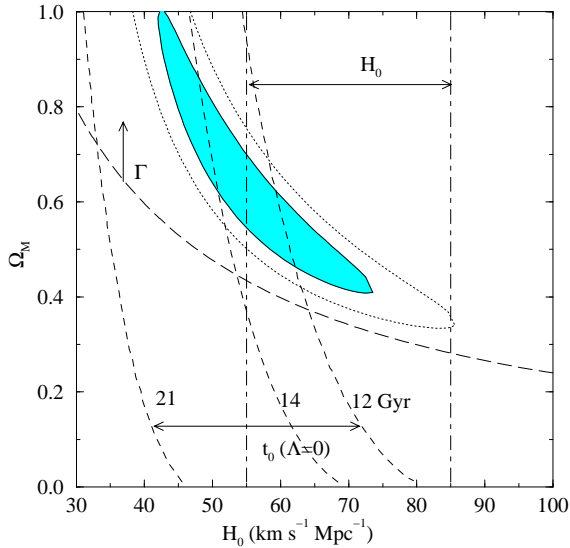} 
\caption{  
68\% (shaded) and 95\% (dotted) confidence regions (``CRs'') in the  
$(H_0, \Omega_{\mathrm{M}})$ plane for CDM models ($\Lambda = 0$) with  
our four standard constraints.  The CRs are closed curves.  Individual  
constraints in this plane are also shown schematically, as explained  
in the text.  
\label{Fig:H-Omega_M_L0}}  
\end{figure}

\begin{figure}[t] 
 
\epsscale{1.0} 
 
\plotone{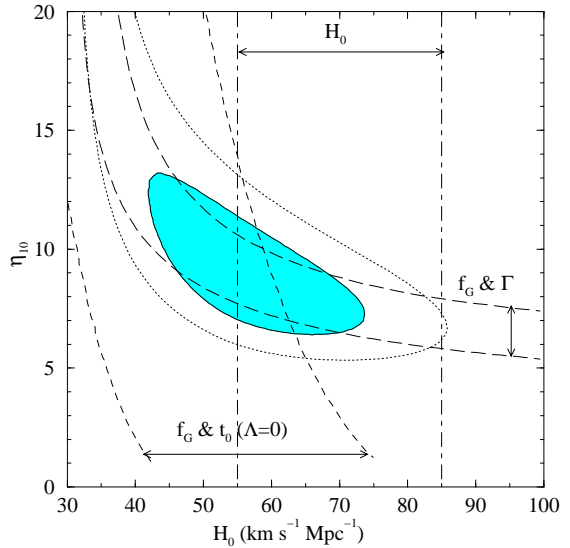} 
\caption{  
Same as Figure~\protect\ref{Fig:H-Omega_M_L0}, but in the $(H_0,  
\eta_{10})$ plane.  Individual and paired constraints are also shown  
schematically.  See text.  
\label{Fig:H-eta_L0}}  
\end{figure}  
 
\begin{figure}[t] 
 
\epsscale{1.0} 
 
\plotone{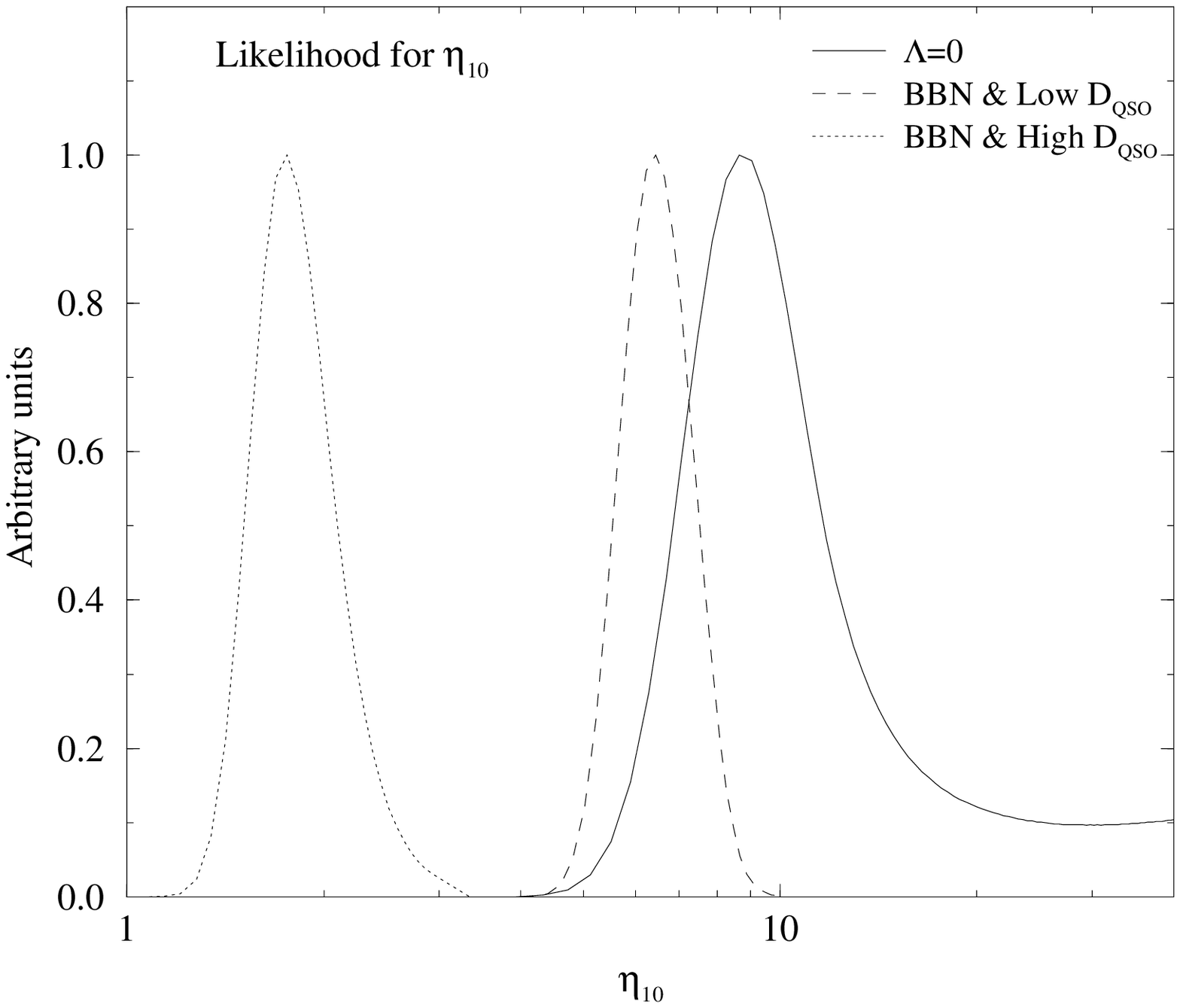} 
\caption{  
Likelihood $\mathcal{L}$$(\eta_{10})$ as a function of $\eta_{10}$   
for CDM models ($\Lambda = 0$) with our four standard constraints (solid).   
Also shown  are the corresponding BBN-predicted likelihoods for the high  
D abundance  (dotted) and for the low D abundance (dashed) inferred from  
QSO absorbers  (cf. Hata et al. 1997).  
\label{Fig:eta-P_L0}}  
\end{figure}

\section{CDM MODELS:  RESULTS}  
\label{Sec:CDM-Models-results}  
  
\subsection{CDM with Standard Constraints}

We begin the presentation of our results by adopting a standard case  
with only four constraints, dropping the $\Omega_{\mathrm{o}}$  
constraint.  For this standard case we assume $n = 1$ and $\Upsilon =  
0.9$, and we apply the following ``standard constraints'':  
$h_{\mathrm{o}} = 0.70 \pm 0.15$, $ t_{\mathrm{o}} = 14^{+7}_{-2}$  
Gyr, $f_{\mathrm{o}} \equiv f_{\mathrm{G}} h^{3/2} = 0.060 \pm 0.006$,  
and $\Gamma_{\mathrm{o}} = 0.255 \pm 0.017$.  Then $\chi^2$ is the sum  
of four terms.  
  
Results for our standard case are displayed in  
Figures~\ref{Fig:H-Omega_M_L0}--\ref{Fig:eta-P_L0}.  
Figures~\ref{Fig:H-Omega_M_L0} and \ref{Fig:H-eta_L0} are a pair which  
can be understood geometrically.  The function $\chi^2 (h,  
\Omega_{\mathrm{M}}, \eta_{10})$ is computed on the three-dimensional  
parameter space.  It has a minimum $\chi^2_{\mathrm{min}}$ in this  
space, which in this case is 1.2 for one degree of freedom (DOF) and  
is located at (0.57, 0.61, 8.7).  This value of  
$\chi^2_{\mathrm{min}}$ is acceptable; it is the 73\% point of the  
distribution.  We may draw a closed surface which encloses this point,  
defined by setting  
\begin{equation}  
\Delta \chi^2 \equiv   
\chi^2(h, \Omega_{\mathrm{M}}, \eta_{10}) - \chi^2_{\mathrm{min}} =  
2.3.  
\label{Eq:Delta-chi2}  
\end{equation}  
The quantity $\Delta\chi^2$ is distributed like a $\chi^2$ variable  
with 3 DOF (Press et al. 1992, Barnett et al. 1996).  Our surface  
$\Delta\chi^2 = 2.3$ is at the 49\% point, so it is a 49\% confidence  
region (``CR'') for the three parameters jointly.  Furthermore, its  
projections on the orthogonal planes ($h, \Omega_{\mathrm{M}}$;  
Fig.~\ref{Fig:H-Omega_M_L0}) and ($h, \eta_{10}$;  
Fig.~\ref{Fig:H-eta_L0}) give the 68\% CRs for the parameters  
pairwise.  These 68\% CRs are shown as closed curves in  
Figures~\ref{Fig:H-Omega_M_L0} and \ref{Fig:H-eta_L0}.  Similarly, we  
construct 95\% CRs in these planes by replacing 2.3 by 6.0 in equation  
(\ref{Eq:Delta-chi2}).  
  
We also show in Figures~\ref{Fig:H-Omega_M_L0} and \ref{Fig:H-eta_L0}  
projected CRs obtained by computing $\chi^2$ for single observables  
alone, or for pairs of observables.  They are drawn by setting  
$\Delta\chi^2 = 1$ and projecting.  These regions are not closed.  
They are merely intended to guide the reader in understanding how  
the various constraints influence the closed contours which show our  
quantitative results.  
  
One-dimensional confidence intervals (CIs) may similarly be  
constructed for any parameter by projecting closed surfaces in  
three-space onto a single axis.  These CIs may be described by a  
likelihood function.  Figure~\ref{Fig:eta-P_L0} shows the likelihood  
function $\mathcal{L}$$(\eta_{10})$ for the parameter $\eta_{10}$.  
Table~\ref{Tab:Table1} shows the one-parameter CIs for the CDM models.

\begin{table*} 
\caption{Best-fit Parameters and Errors for Four Standard Constraints 
\label{Tab:Table1}}  
\vspace{1ex} 
\begin{center} 
\begin{tabular}{l l l l l} 
\tableline 
\tableline 
Parameter   &    
	\multicolumn{2}{c}{$\Lambda$ = 0, CDM Model}      
                           & \multicolumn{2}{c}{$k$ = 0, $\Lambda$CDM Model}\\  
\tableline 
$\eta_{10}$ & $8.7\,^{+2.3}_{-1.6}$ & $(>6.1)$&  
                                         $8.4 \, ^{+2.1}_{-1.5}$ & $(>5.8)$\\ 
$\Omega_{\mathrm{B}} $ & $0.10\,^{+0.08}_{-0.04}$ & $(>0.04)$  
                                    & $0.08 \, ^{+0.06}_{-0.03}$ & $(>0.03)$\\ 
$\Omega_{\mathrm{M}}  $& $0.61\,^{+0.20}_{-0.14}$ & $(>0.39)$  
                                     & $0.53\,^{+0.19}_{-0.11}$ & $(>0.35)$\\ 
$H_0$ (km s$^{-1}$ Mpc$^{-1}$) & $57\,^{+11}_{-10}$ & $(36 - 80)$  
                                        & $62\,^{+ 13}_{-11}$ & $(39 - 87)$\\ 
$t_0$ (Gyr) & $12.6\,^{+1.9}_{-1.6}$ & $(9.7 - 16.6)$  
                                 & $12.9\,^{+1.5}_{-1.4}$ & $(10.5 - 16.1)$\\ 
\tableline 
\end{tabular} 
\end{center} 
 
\tablenotetext{}{Error bars are for 68\% CI; range in parentheses is for 95\%.} 
 
\end{table*}

\subsection{CDM:  Discussion}

It is well known that the condition $\Omega_{\mathrm{M}} h \approx  
0.25$ poses some threat to the SCDM $(\Omega_{\mathrm{M}} = 1)$ model.  
Figure~\ref{Fig:H-Omega_M_L0} shows that this threat is far from  
acute, with our more accurate form of the $\Gamma$ constraint given in  
equation~(\ref{Eq:Gamma_th}), as long as the error on $h_{\mathrm{o}}$  
is large (0.15) and BBN constraints are discarded.  (Note again that  
we have not applied the constraint $\Omega_{\mathrm{o}} = 0.2 \pm  
0.1$.)  Even our 68\% contour extends to $\Omega_{\mathrm{M}}= 1$.  
With the large error bar, the corresponding value of $H_{0}$, $h =  
0.43$, is accepted.  The exponential term in equation~(\ref{Eq:Gamma_th})  
becomes significant because the $f_{\mathrm{G}}$ constraint forces  
$\Omega_{\mathrm{B}}$ to increase with $\Omega_{\mathrm{M}}$, allowing  
the product $\Omega_{\mathrm{M}}h$ to exceed 0.25.  This has been noted  
before (White et al. 1996, Lineweaver et al. 1997).

\begin{figure}[t] 
 
\epsscale{1.0} 
 
\plotone{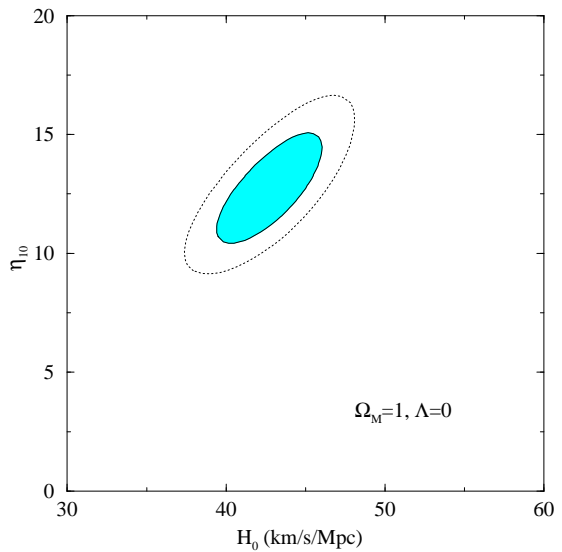} 
\caption{ 
68\% (shaded) and 95\% (dotted) CRs in the $(H_0, \eta_{10})$  
plane for CDM models ($\Lambda = 0$) with our four standard constraints,  
but with  $\Omega_{\mathrm{M}}$ now fixed at unity (SCDM models).  
\label{Fig:H-eta_OM1}}  
\end{figure}

We have tested the SCDM model by fixing $\Omega_{\mathrm{M}}$ at unity  
and fitting the four standard constraints with the remaining two  
parameters.  The CRs for $\eta_{10} - H_0$ are shown in  
Figure~\ref{Fig:H-eta_OM1}.  We find $\chi^2_{\mathrm{min}} = 3.4$ for  
2 DOF (82\% CL), which is acceptable.  However, this case encounters  
severe problems since only $h < 0.48$ and $\eta_{10} > 8$ are accepted  
at the 95\% contour.  Indeed, $\eta_{10} \gtrsim 8$ if $h \approx 0.4$  
and $\eta_{10} \approx 15$ if $h \approx 0.48$.  Such large $\eta$  
values pose a serious threat to the consistency between the  
predictions of BBN and the primordial abundances of the light elements  
inferred from observations (e.g., Hata et al. 1996).  Indeed, this  
``solution" is only acceptable because of our very generous error bar  
for $h$ and because we have discarded BBN constraints.

\begin{figure}[t] 
 
\epsscale{1.0} 
 
\plotone{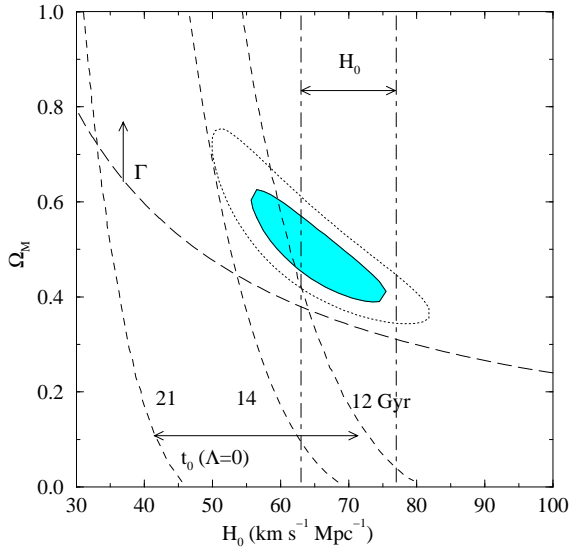} 
\caption{ 
Same as Figure~\protect\ref{Fig:H-Omega_M_L0}, but with the constraint  
$h_{\mathrm{o}} = 0.70 \pm 0.15$ replaced by $h_{\mathrm{o}} = 0.70  
\pm 0.07.$  
\label{Fig:H-Omega_M_L0_H70+-7}}  
\end{figure}  
 
\begin{figure}[t] 
 
\epsscale{1.0} 
 
\plotone{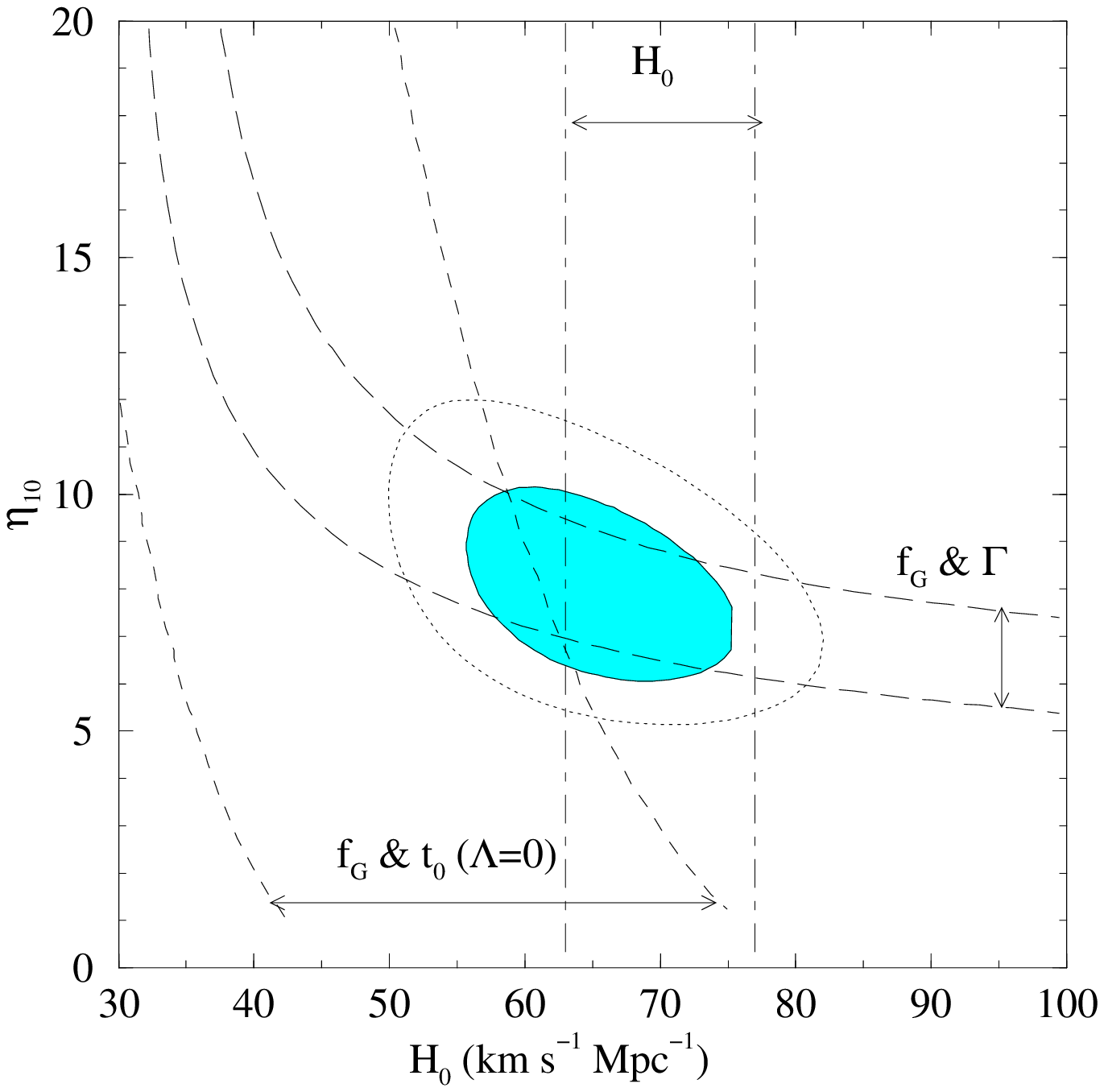} 
\caption{  
Same as Figure~\protect\ref{Fig:H-eta_L0}, but with the constraint $  
h_{\mathrm{o}} = 0.70 \pm 0.15$ replaced by $h_{\mathrm{o}} = 0.70  
\pm 0.07.$  
\label{Fig:H-eta_L0_H70+-7}}  
\end{figure}  
 
\begin{figure}[t] 
 
\epsscale{1.0} 
 
\plotone{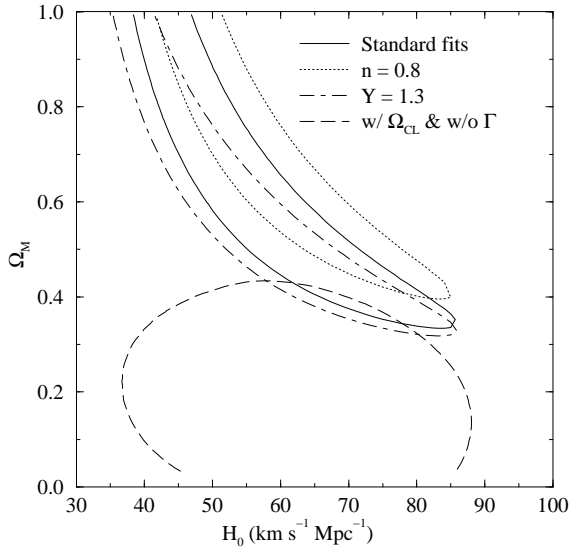} 
\caption{  
Same as Figure~\protect\ref{Fig:H-Omega_M_L0} for variations on our  
CDM results ($\Lambda = 0$) for four standard constraints.  The variations  
are taken one at a time.  Only 95\% CRs are shown.  The result from  
Figure~\protect\ref{Fig:H-Omega_M_L0} (no variation; solid curve);  
``red'' tilt $n = 0.8$ (instead of $n=1$; dotted curve); gas  
enhancement factor $\Upsilon = 1.3$ (instead of 0.9; dot-dash curve);  
without the $\Gamma$ constraint {\it and} with the cluster constraint  
$\Omega_{\mathrm{o}} = 0.2 \pm 0.1$ (dashed curve).  
\label{Fig:H-Omega_M_L0_var}}  
\end{figure}

\begin{figure}[t] 
 
\epsscale{1.0} 
 
\plotone{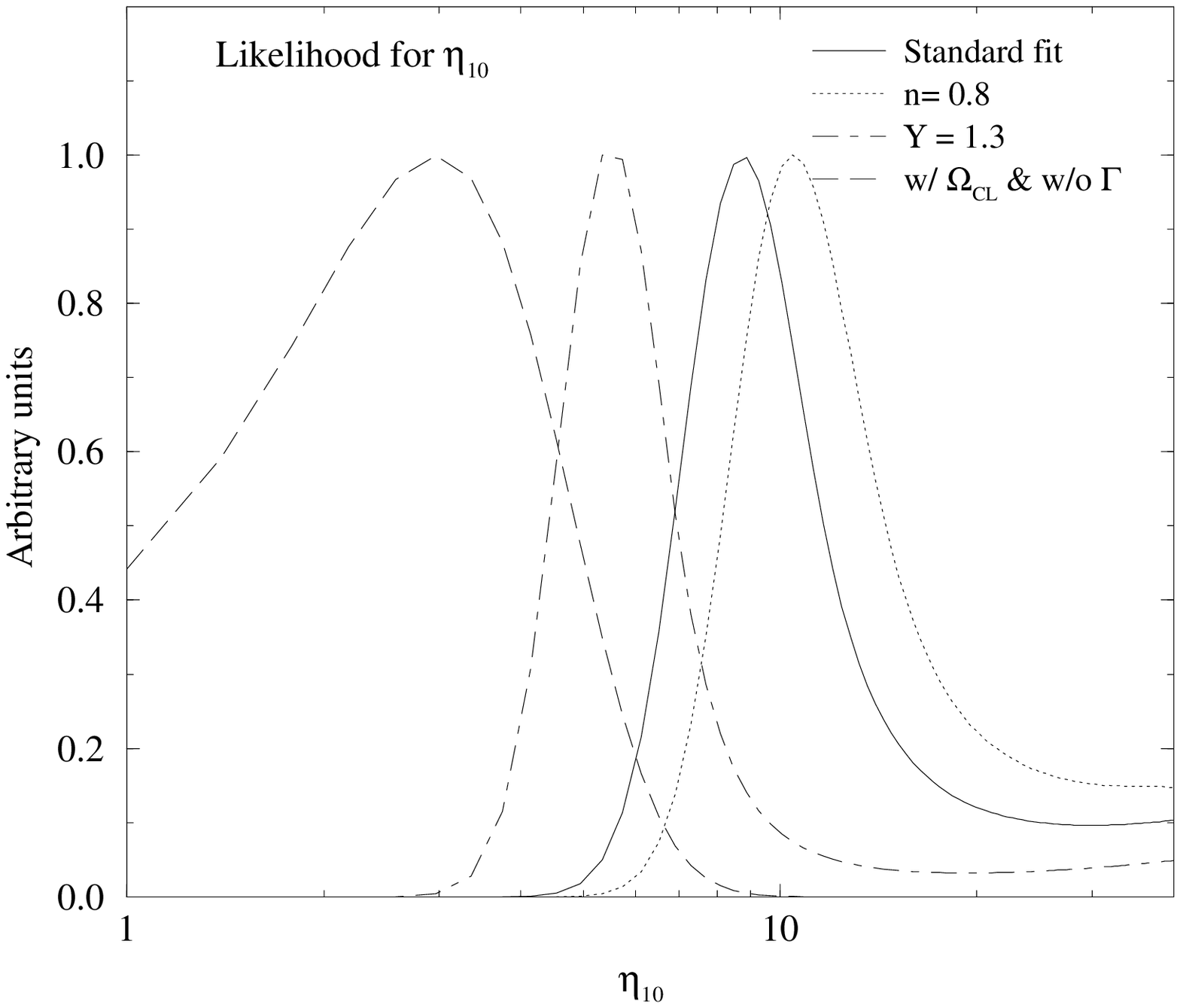} 
\caption{  
Likelihoods $\mathcal{L}$$(\eta_{10})$ as a function of $\eta_{10}$  
for the CDM models ($\Lambda = 0$) with the same variations as in  
Figure~\protect\ref{Fig:H-Omega_M_L0_var}.  
\label{Fig:eta-P_L0_var}}  
\end{figure}

When $h$ is better known, the situation for SCDM will change.  As an  
illustration, in Figure~\ref{Fig:H-Omega_M_L0_H70+-7} we return to our  
three standard variables but replace our standard constraint on $H_0$  
with $h_{\mathrm{o}} = 0.70 \pm 0.07$, assuming, arbitrarily, a 10\%  
error.  The $\chi^2_{\mathrm{min}}$ is now 2.2 for 1 DOF (86\% CL), so  
we can still accept the basic Friedmann model, but SCDM is now  
excluded strongly.  In this case the corresponding allowed range of  
$\eta$, shown in Figure~\ref{Fig:H-eta_L0_H70+-7}, is not in strong  
conflict with BBN although the predicted $^4$He abundance is larger  
than that inferred from observations of extragalactic \hii regions  
(OS, OSS).  
  
Returning to our CDM case with standard constraints, it is less well  
known that the $\Gamma$ constraint also poses a threat to {\it  
low}-density models (Liddle et al. 1996, Kolatt \& Dekel 1997, White  
\& Silk 1996).  From Figure~\ref{Fig:H-Omega_M_L0} it is apparent that  
there is tension between our four standard constraints and  
$\Omega_{\mathrm{o}} = 0.2 \pm 0.1$.  Since the 95\% contour does not  
even extend downward to $\Omega_{\mathrm{M}} = 0.3$, we refrained from  
using this cluster constraint.  The $t_{\mathrm{o}}$ and  
$\Gamma_{\mathrm{o}}$ constraints, combined, force the parameters  
upward out of the lower part of the figure, and favor  
$\Omega_{\mathrm{M}} \gtrsim 0.4$.  We could nevertheless force a fit  
to all five constraints and draw CRs.  We have done this, and we find  
that $\chi^2_{\mathrm{min}}$ is 7.8 for 2 DOF (98\% CL); i.e., we  
reject the combined fit with 98\% confidence.  
  
Among our principal results is that our standard CDM case favors large  
values of the baryon-to-photon ratio, $\eta_{10} = 8.7^{+2.3}_{-1.6}$  
(see Figure~\ref{Fig:eta-P_L0} and Table~\ref{Tab:Table1}).  It is the  
$f_{\mathrm{o}}$ and $\Gamma_{\mathrm{o}}$ constraints which,  
together, force us to large $\eta_{10}$.  Also shown in  
Figure~\ref{Fig:eta-P_L0} are the likelihoods for $\eta$ derived in  
Hata et al. (1997) for the high deuterium abundance inferred for some  
QSO absorbers (Songaila et al.  1994, Carswell et al. 1994, Rugers \&  
Hogan 1996) and for the lower D abundance inferred for others (Tytler  
et al. 1996, Burles \& Tytler 1996).  It is clear from  
Figure~\ref{Fig:eta-P_L0} that our results here favor the high-$\eta$,  
low-D choice which is consistent with local deuterium observations  
(Linsky et al. 1993) and Galactic chemical evolution (Steigman \& Tosi  
1992, 1995, Edmunds 1994, Tosi 1996).  BBN consistency with the  
observed lithium abundances in very metal-poor halo stars requires  
that these stars have reduced their surface lithium abundance by a  
modest factor, $\lesssim 2-3$.  However, for consistency with standard  
BBN predictions for helium, our high value for $\eta$ requires that  
the abundances inferred from the low-metallicity, extragalactic \hii  
regions are systematically biased low.  This high-$\eta$ range is  
consistent with estimates of the baryon density derived from  
observations of the Ly-$\alpha$ forest (Hernquist et al.  1996,  
Miralda-Escud\'e et al. 1996, Rauch et al. 1997, Croft et al. 1997,  
Weinberg et al. 1997, Bi \& Davidsen 1997).

\subsection{CDM:  More Variations}  
  
Because we dropped the cluster-determined constraint  
$\Omega_{\mathrm{o}} = 0.2 \pm 0.1$, it is of interest to see how the  
CRs in Figure~\ref{Fig:H-Omega_M_L0} are affected if we apply instead  
an alternative constraint to $\Omega_{\mathrm{M}}$.  If, for example,  
we adopt the Dekel-Rees constraint, $\Omega_{\mathrm{DR}}$ [equation  
(\ref{Eq:Omega_DR})], which implies a substantial contribution to  
$\Omega_{\mathrm{M}}$ arising from mass not traced by light, this does  
not change Figure~\ref{Fig:H-Omega_M_L0} by much because  
$\Omega_{\mathrm{M}} > 0.4$ was favored already by our four standard  
constraints.  Small $h$ and large $\Omega_{\mathrm{M}}$ are now  
favored slightly more.  Because this makes little difference, we will  
proceed in most cases without any constraint $\Omega_{\mathrm{M}}$.  
The consequences for $\eta$ are found in Table~\ref{Tab:Table2}.

\begin{table} 
\caption{Variations: Best-fit Values and Errors for $\eta_{10}$ 
\label{Tab:Table2}} 
\vspace{1ex} 
\begin{center} 
\begin{tabular}{l l l} 
\tableline 
\tableline 
Variation     & \multicolumn{2}{c}{$\eta_{10}$}                        \\ 
\tableline 
With $\Omega_{\mathrm{DR}}$ & $9.2\;^{+2.2}_{-1.5}$  & $(>6.5)$\\ 
``Red'' tilt $n = 0.8$      & $10.8\;^{+3.5}_{-2.0}$ & $(>7.3)$\\ 
Positive gas bias $\Upsilon = 1.3$ & $5.7\;^{+1.2}_{-0.9}$ & $(>4.0)$\\ 
Without $\Gamma$; With $\Omega_{\mathrm{CL}}$ & $3.1 \pm 1.6$ & $(<6.5)$\\ 
$\Gamma = 0.25 \pm 0.05$    & $8.2\;^{+3.2}_{-2.2}$ & $(>4.2)$\\ 
$\Gamma = 0.15 \pm 0.04$    & $4.6\;^{+1.9}_{-1.2}$ & $(2.1 - 9.2)$\\  
\tableline 
\end{tabular} 
 
\tablenotetext{}{Error bars are for 68\% CI and range in parentheses  
is for 95\%.} 
 
\end{center} 
\end{table}

Figure~\ref{Fig:H-Omega_M_L0_var}, the analog of  
Figure~\ref{Fig:H-Omega_M_L0}, shows the effects of some other  
variations, taken one at a time.  Here we consider only $\Lambda = 0$  
models and show only the 95\% CRs.  The corresponding likelihoods for  
$\eta$ are shown in Figure~\ref{Fig:eta-P_L0_var}.  Tilt in the  
primordial spectrum, for example, has been investigated in many papers  
(Liddle et al. 1996a,b, White et al. 1996, Kolatt \& Dekel 1997, White  
\& Silk 1996, Liddle \& Viana 1996).  We show the effect of a moderate  
``red tilt'' ($n = 0.8$ instead of $n = 1$).  The  
$\chi^2_{\mathrm{min}}$ value is 1.5 for 1 DOF (78\% CL).  The favored  
likelihood range for $\eta_{10}$ is now higher, though $\eta_{10}  
\approx 7$ is still allowed (see Figure~\ref{Fig:eta-P_L0_var}).  With  
this tilt the $\Gamma$ constraint favors higher $\Omega_{\mathrm{M}}$, 
so that the SCDM model is allowed for $h$ up to nearly 0.5.  However, 
as can be seen in Table~\ref{Tab:Table2}, the higher allowed range for 
$\eta$ threatens the consistency of BBN.  Conversely, a ``blue'' tilt, 
$n > 1$ (Hancock et al. 1994), would move the CR downward and allow 
models with $\Omega_{\mathrm{M}} \le 0.3$ at high $h$. 
  
Changing to a gas enhancement factor $\Upsilon = 1.3$ (a modest 
positive enhancement of gas in clusters) instead of 0.9 does not 
change the contours in Figure~\ref{Fig:H-Omega_M_L0} by much, 
particularly at the low-$\Omega_{\mathrm{M}}$ end, where the 
exponential term in $\Gamma$ is close to unity. 
The $\chi^2_{\mathrm{min}}$  
value for this case is 1.1 for 1 DOF (71\% CL).  Although the  
acceptable range for $\eta_{10}$ moves downward (see  
Figure~\ref{Fig:eta-P_L0_var}), $\eta_{10} \le 4$ is still excluded,  
disfavoring the low D abundance inferred from some QSO absorbers and  
favoring a higher helium abundance than is revealed by the \hii region  
data.  In \S 2.6 we mentioned the possibility that the fraction of  
cluster mass in baryons in galaxies, isolated stars, and machos  
$(f_{\mathrm{GAL}})$ might be larger -- even much larger -- than is  
implied by equation~(\ref{Eq:f_G/f_GAL}).  
Equations~(\ref{Eq:f_G/f_GAL}) and (\ref{Eq:f_Gh}) show that a large  
$f_{\mathrm{GAL}}$ would affect the CRs in much the same way as a {\it  
small} $\Upsilon$, favoring even higher values of  
$\Omega_{\mathrm{M}}$ and $\eta_{10}$.  
  
The $\Gamma$ constraint is crucial for our standard results favoring  
high $\Omega_{\mathrm{M}}$ and high $\eta$.  If, for example, we drop  
the $\Gamma$ constraint and in its place use the cluster estimate  
$\Omega_{\mathrm{o}} = 0.2 \pm 0.1$, low $\Omega_{\mathrm{M}}$ and  
low $\eta$ are now favored (see Figures~\ref{Fig:H-Omega_M_L0_var} and  
\ref{Fig:eta-P_L0_var}).

\begin{figure}[t] 
 
\epsscale{1.0} 
 
\plotone{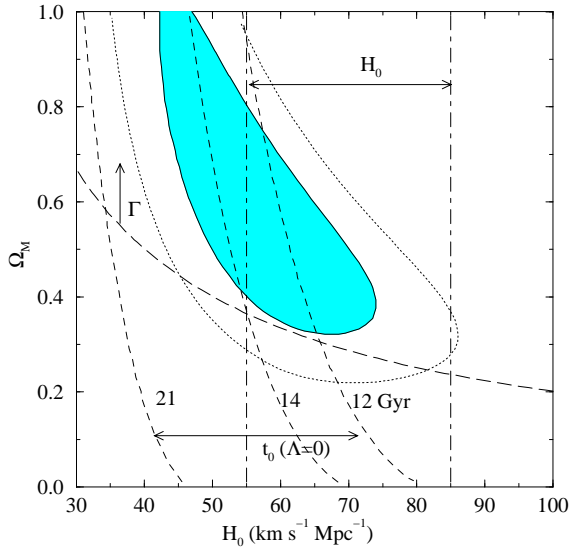} 
\caption{  
Same as Figure~\protect\ref{Fig:H-Omega_M_L0}, but with the  
shape-parameter constraint $\Gamma_{\mathrm{o}} = 0.25 \pm 0.05$  
(instead of 0.255 $\pm$ 0.017).  
\label{Fig:H-Omega_M_L0_G+-0.05}}  
\end{figure}  
 
\begin{figure}[t] 
 
\epsscale{1.0} 
 
\plotone{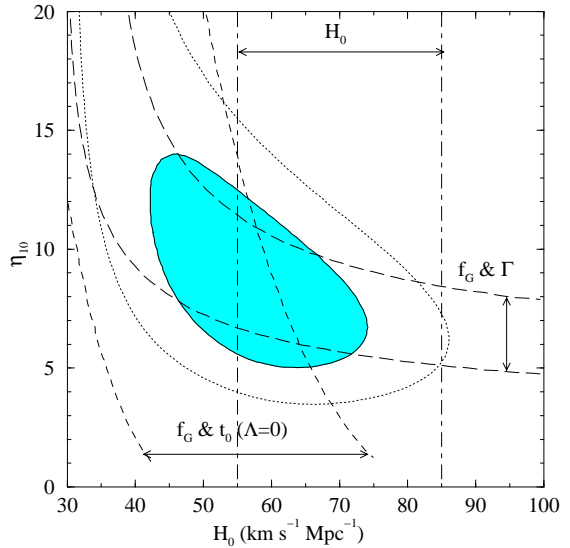} 
\caption{  
Same as Figure~\protect\ref{Fig:H-eta_L0}, but with the  
shape-parameter constraint $\Gamma_{\mathrm{o}} = 0.25 \pm 0.05$  
(instead of 0.255 $\pm$ 0.017).  
\label{Fig:H-eta_L0_G+-0.05}}  
\end{figure}

Earlier we mentioned that the Peacock and Dodds (1994) estimate of  
$\Gamma_{\mathrm{o}}$ may have unrealistically small error bars.  
Given that the shape parameter plays such an important role in our  
analysis, we have considered the effects of relaxing the uncertainty  
in $\Gamma_{\mathrm{o}}$.  In Figures~\ref{Fig:H-Omega_M_L0_G+-0.05}  
and \ref{Fig:H-eta_L0_G+-0.05}, the analogs of  
Figures~\ref{Fig:H-Omega_M_L0} and \ref{Fig:H-eta_L0}, we show our  
results for $\Gamma_{\mathrm{o}}$ = 0.25 $\pm$ 0.05.  As expected our  
CRs have expanded and the best-fit values of $\Omega_{\mathrm{M}}$,  
$h$ and $\eta_{10}$ have shifted: $\Omega_{\mathrm{M}} =  
0.48^{+0.22}_{-0.15}$, $h = 0.58 \pm 0.22$ and $\eta_{10} =  
8.2^{+3.2}_{-2.2}$.  Now the SCDM model with $\Omega_{\mathrm{M}}$ =  
1, $h = 0.45$, and $\eta_{10} = 13$ is acceptable (80\%).  Although  
the uncertainties are larger, low $\eta_{10}$ is still disfavored.  If  
we add the Dekel-Rees estimate of $\Omega_{\mathrm{M}}$, the  
five-constraint fit favors somewhat higher values of  
$\Omega_{\mathrm{M}}$ and $\eta_{10}$ and slightly lower values of  
$h$.  In contrast, if instead we include the cluster estimate, we find  
a barely acceptable fit ($\chi^2_{\mathrm{min}}$ = 5.0 for 2 DOF, 92\%  
CL), which favors lower values of $\Omega_{\mathrm{M}}$ and  
$\eta_{10}$ and slightly higher values of $h$.  
 
\subsection{CDM: A Smaller Shape Parameter} 
 
The results above show that the Peacock-Dodds shape parameter  
$\Gamma_{\mathrm{o}} \approx 0.255$, which we have used, clashes with 
$\Omega_{\mathrm{M}} \approx 0.2$ (the estimate from clusters). The  
agreement does not become good even if error bars $\pm 0.05$ on 
$\Gamma_{\mathrm{o}}$ are assumed. 
 
A new determination of $\Gamma_{\mathrm{o}}$ from the IRAS redshift survey  
(Webster et al. 1998) gives $\Gamma_{\mathrm{o}} = 0.15 \pm 0.08$ at 95\%  
confidence.  Assuming that the statistics are roughly Gaussian, we can  
represent this at $\pm 1\sigma$ as 
\begin{equation}  
\Gamma_{\mathrm{o}} = 0.15 \pm 0.04.  
\label{Eq:newGamma_obs}  
\end{equation}  
Combining equation (16) and equation (12) (used earlier) in quadrature would 
give $\Gamma_{\mathrm{o}} = 0.239 \pm 0.016$ -- very close to equation (12).  
Throwing in equation (14) would bring us even closer to equation (12), which  
dominates because of its small error.  Combining the estimates would be unwise,  
because they do not agree well. 
         
This small value from the IRAS survey is not entirely new (cf. Fisher,  
Scharf, \& Lahav 1994).  It has received little attention because the  
larger value, $\Gamma_{\mathrm{o}} \approx 0.25$, was already seen as  
a major challenge to the popular SCDM model ($\Omega_{\mathrm{M}} = 1$).  
The smaller value poses an even more severe challenge to the SCDM model.   
But it gives more scope to low-density models, which are popular now.   
Until the discrepant values of $\Gamma_{\mathrm{o}}$ are understood,  
we think it wise to show joint CRs using separately the larger and the  
smaller values of $\Gamma_{\mathrm{o}}$.        
 
\begin{figure}[t] 
 
\epsscale{1.043} 
 
\plotone{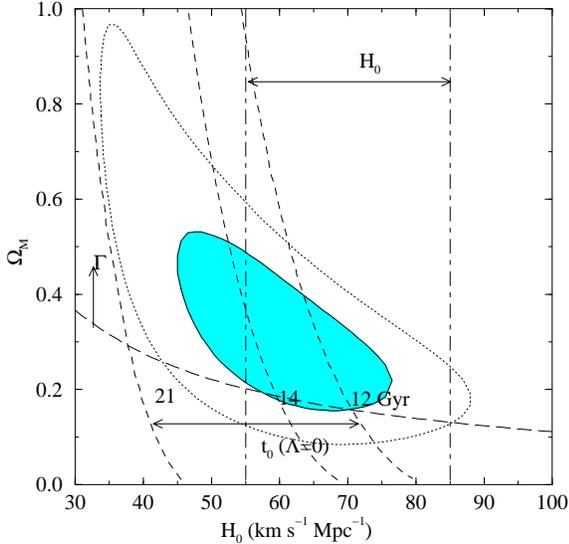} 
\caption{ Same as Figure~\protect\ref{Fig:H-Omega_M_L0}, but with the  
shape-parameter constraint $\Gamma_{\mathrm{o}} = 0.15 \pm 0.04$ (instead  
of $0.255 \pm 0.017$). 
\label{Fig:H-Omega_M_L0_G0.15}}  
\end{figure}

\begin{figure}[t] 
 
\epsscale{1.043} 
 
\plotone{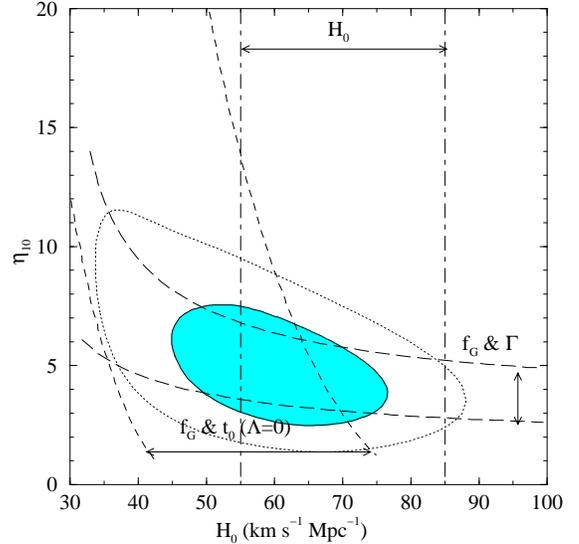} 
\caption{ Same as Figure~\protect\ref{Fig:H-eta_L0}, but with the  
shape-parameter constraint $\Gamma_{\mathrm{o}} = 0.15 \pm 0.04$  
(instead of $0.255 \pm 0.017$). 
\label{Fig:H-eta_L0_G0.15}}  
\end{figure}

Figures~\ref{Fig:H-Omega_M_L0_G0.15} and \ref{Fig:H-eta_L0_G0.15}, 
analogs of Figures~\ref{Fig:H-Omega_M_L0} and \ref{Fig:H-eta_L0}, show 
CRs for our four standard constraints, but with $\Gamma_{\mathrm{o}} = 
0.255\pm 0.017$ replaced by $\Gamma_{\mathrm{o}} = 0.15 \pm 0.04$. The 
$\chi^2_{\mathrm{min}}$ is 0.63 for 1 DOF (good) and is located at 
($h$, $\Omega_{\mathrm{M}}$, $\eta_{10}$) = (0.60, 0.30, 4.6).

The CRs now exclude the SCDM model strongly and favor low density. The 
value $\eta_{10} \approx 5$, favored by the Burles and Tytler 
(1997a,b,c) deuterium abundance determination (see \S 1), is now near 
the point of optimum fit.  The CRs clearly would accept the added 
cluster constraint $\Omega_{\mathrm{M}} \approx 0.2$ if we were to 
apply it.  But note that in our three-dimensional CRs, low 
$\Omega_{\mathrm{M}}$ goes with low $\eta_{10}$, because of the 
$f_{\mathrm{G}}$ constraint.  For example, the combinations (0.7, 0.2, 
3) and (0.7, 0.3, 5) give good fits, while (0.7, 0.2, 5) and (0.7, 
0.3, 3) give poor fits.  In general, the CRs give some preference to 
$\eta_{10} \approx 5$.  But even a value as small as $\eta_{10} 
\approx 2$ now lies within the 95\% CI and cannot be excluded without 
BBN evidence (see Table~\ref{Tab:Table2}).

\section{$\Lambda$CDM MODELS:  RESULTS}  
\label{Sec:Lambda-CDM-Models}

\begin{figure}[t] 
 
\epsscale{1.0} 
 
\plotone{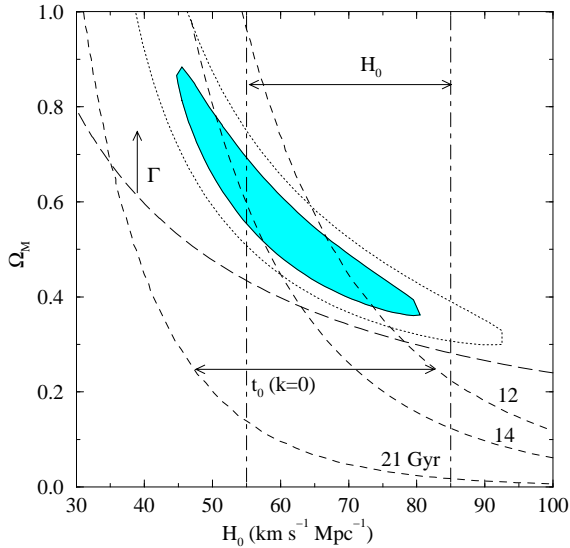} 
\caption{  
Same as Figure~\protect\ref{Fig:H-Omega_M_L0}, but for  
$\Lambda$CDM (flat) models.  
\label{Fig:H-Omega_M_k0}}  
\end{figure}

\begin{figure}[t] 
 
\epsscale{1.0} 
 
\plotone{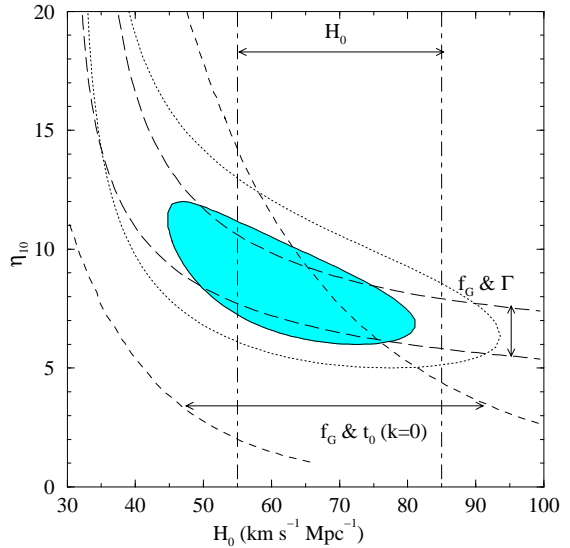} 
\caption{  
Same as Figure~\protect\ref{Fig:H-eta_L0}, but for  
$\Lambda$CDM (flat) models.  
\label{Fig:H-eta_k0}}  
\end{figure}

Turning to models with nonzero $\Lambda$, we consider here only the  
popular flat ($k$ = 0) ``$\Lambda$CDM'' models with $\Omega_\Lambda =  
1 - \Omega_{\mathrm{M}}$, where $\Omega_\Lambda \equiv \Lambda  
/(3H_0^2)$.  This means that there are still only three free  
parameters.  The five constraints discussed in \S 2 are still in  
force, except that the product of the age and the Hubble parameter is  
a different function of $\Omega_{\mathrm{M}} = 1 - \Omega_\Lambda$: $  
t = 9.78\,h^{-1}f(\Omega_{\mathrm{M}};k=0)$ Gyr [Carroll, Press, \&  
Turner 1992, equation (17)].  For a given $\Omega_{\mathrm{M}} < 1$,  
the age is longer for the flat $(k = 0)$ model than for the $\Lambda =  
0$ model.  Figures~\ref{Fig:H-Omega_M_k0} and \ref{Fig:H-eta_k0} show  
the results for our four standard constraints, with no direct  
$\Omega_{\mathrm{M}}$ constraint.  Figures~\ref{Fig:H-Omega_M_k0} and  
\ref{Fig:H-eta_k0} differ very little from  
Figures~\ref{Fig:H-Omega_M_L0} and \ref{Fig:H-eta_L0}, of which they 
are the analogs.  The longer ages do allow the CRs to slide farther 
down toward large $h$ and small $\Omega_{\mathrm{M}}$.  The 
$\chi^2_{\mathrm{min}}$ is 0.8 for 1 DOF (good).  One-dimensional CIs 
are listed in Column 3 of Table~\ref{Tab:Table1}.  Because of the 
longer ages at low $\Omega_{\mathrm{M}}$ (high $\Omega_{\Lambda}$), we 
can now accept $\Omega_{\mathrm{o}}$ as a fifth constraint [although 
we remind the reader that the constraint $\Omega \approx 0.2$ may not 
be appropriate to a $\Lambda$CDM model (\S 2.5)]; the 
$\chi^2_{\mathrm{min}}$ is 5.4 for 2 DOF (93\%, barely acceptable). 
In this case large $\Omega_{\mathrm{M}}$ and small $h$ are now 
excluded while $\eta_{10} > 4$ is still favored strongly.

We have not imposed any direct constraint on $\Omega_\Lambda$. There 
are claims that, for a $\Lambda$CDM model, $\Omega_\Lambda < 0.51$ 
(based on limited statistics of seven supernovae; Perlmutter et 
al. 1997) and $\Omega_\Lambda < 0.66$ (based on a paucity of lensing 
events; Kochanek 1996). The lensing constraint has been in dispute 
because of absorption, but recent work indicates that absorption is 
probably unimportant (Kochanek 1996; Falco, Kochanek, \& Munoz 
1997). These $\Omega_\Lambda$ constraints agree in a general way with 
our result $\Omega_{\mathrm{M}} \gtrsim 0.4$ 
(Fig.~\ref{Fig:H-Omega_M_k0}).

\section{CONCLUSIONS}  
\label{Sec:Conclusions}  
  
If BBN constraints on the baryon density are removed (or relaxed), the  
interaction among the shape-parameter $(\Gamma)$ constraint, the  
$f_{\mathrm{G}}$ (cosmic baryon fraction) constraint, and the value of  
$\eta_{10}$ assumes critical importance.  These constraints still  
permit a flat CDM model, but only as long as $h < 0.5$ is allowed by  
observations of $h$.  The $f_{\mathrm{G}}$ constraint means that large  
$\Omega_{\mathrm{M}}$ implies fairly large $\Omega_{\mathrm{B}}$.  
Therefore the exponential term in $\Gamma$ becomes important, allowing  
$\Omega_{\mathrm{M}} = 1$ to satisfy the $\Gamma$ constraint.  Values  
of $\eta_{10} \approx 8-15$ are required (Fig.~\ref{Fig:H-eta_OM1}).  
The best-fit SCDM model has $h \approx 0.43$ and $\eta_{10} \approx  
13$, which is grossly inconsistent with the predictions of BBN and the  
observed abundances of D, $^4$He, and $^7$Li.  For $h > 0.5$ a fit to  
SCDM is no longer feasible (Fig.~\ref{Fig:H-eta_OM1}).  The SCDM model  
is severely challenged.  
  
The $\Gamma$ and age constraints also challenge low-density CDM  
models.  The $\Gamma$ constraint permits $\Omega_{\mathrm{M}} < 0.4$  
only for high $h$, while the age constraint forbids high $h$, so  
$\Omega_{\mathrm{M}} \gtrsim 0.4$ is required.  Values $\eta_{10}  
\gtrsim 6$ are favored strongly over $\eta_{10} \lesssim 2$.  The  
bound $\Omega_{\mathrm{M}} \gtrsim 0.4$ conflicts with the added 
cluster constraint $\Omega_{\mathrm{o}} = 0.2 \pm 0.1$ at the 98\% CL, 
suggesting strongly that there is additional mass not traced by light. 
  
Although a few plausible variations on the CDM models do not affect  
the constraints very much (Figs.~\ref{Fig:H-Omega_M_L0_var} --  
\ref{Fig:H-eta_L0_G+-0.05}), 
removing the $\Gamma$ constraint would have a dramatic effect.  Both 
high and low values of $\Omega_{\mathrm{M}}$ would then be permitted. 
Adopting a smaller observed $\Gamma_{\mathrm{o}} \approx 0.15$ from 
the IRAS redshift survey also makes a difference.  Values 
$\Omega_{\mathrm{M}} \approx 0.3$ and $\eta_{10} \approx 5$ are then 
favored, but even $\eta_{10} 
\approx 2$ is not excluded. 
  
At either low or high density, the situation remains about the same  
for the $\Lambda$CDM models (Figs.~\ref{Fig:H-Omega_M_k0} \&  
\ref{Fig:H-eta_k0}).  Because the ages are longer, we can tolerate  
$\Omega_{\mathrm{M}} \approx 0.3$ for $h = 0.85$.  The $\Lambda$CDM 
model therefore accepts (barely) the added constraint 
$\Omega_{\mathrm{o}} = 0.2 \pm 0.1$ at the 7\% CL, even with the 
larger $\Gamma_{\mathrm{o}} \approx 0.255$.  Improved future 
constraints on $\Omega_{\Lambda}$ will come into play here. 
  
Having bounded the baryon density using data independent of 
constraints from BBN, we may explore the consequences for the light 
element abundances.  In general, our fits favor large values of 
$\eta_{10}$ ($\gtrsim 5$) over small values ($\lesssim 2$).  While 
such large values of the baryon density are consistent with estimates 
from the Ly-$\alpha$ forest, they may create some tension for BBN. 
For deuterium there is no problem, since for $\eta_{10} \gtrsim 5$ the 
BBN-predicted abundance, (D/H)$_{\mathrm{P}} \lesssim 4 \times 
10^{-5}$ (2$\sigma$), is entirely consistent with the low abundance 
inferred for some of the observed QSO absorbers (Tytler et al. 1996; 
Burles \& Tytler 1996; Burles \& Tytler 1997a,b,c).  Similarly, the 
BBN-predicted lithium abundance, (Li/H)$_{\mathrm{P}} \gtrsim 1.7 
\times 10^{-10}$, is consistent with the observed surface lithium 
abundances in the old, metal-poor stars (including, perhaps, some 
minimal destruction or dilution of the prestellar lithium).  However, 
the real challenge comes from $^4$He where the BBN prediction for 
$\eta_{10} \gtrsim 5$, Y$_{\mathrm{P}} 
\gtrsim 0.246$ (2$\sigma$), is to be contrasted with the \hii region 
data which suggest Y$_{\mathrm{P}} \lesssim 0.238$ (OS, OSS).

\acknowledgments

We wish to thank Neta Bahcall, Rupert Croft, Eli Dwek, Gus Evrard, 
Brian Fields, Andrew Gould, David Graff, Craig Hogan, John Huchra, 
Garth Illingworth, Sasha Kashlinsky, Chris Kochanek, Paul Langacker, 
Andrew Liddle, Rich Mushotzky, John Peacock, Martin Rees, Caleb 
Scharf, Allen Sweigart, and David Weinberg for helpful advice. 
J.E.F. and G.S.  worked on this paper during several workshops at the 
Aspen Center for Physics.  The research of G.S. at Ohio State is 
supported by DOE grant DE-FG02-91ER-40690.  N.H. is supported by the 
National Science Foundation Contract No. NSF PHY-9513835. 
  
\vspace{1ex}  
  
While preparing the final version of our manuscript, we saw the paper 
by Lineweaver and Barbosa (1998), which has some overlap with our 
work.  Their analysis in \S 4 has some similarities to our 
calculations, but there are some important differences.  For example, 
they have only two free parameters since $\eta_{10}$ is incorporated 
by a questionable procedure relying on BBN and is not free, and they 
use a different theoretical expression and different error bars for 
the shape parameter $\Gamma$.  Their \S\S 1-3 are of more interest, 
since they apply entirely different constraints from the CMB angular 
power spectrum.  It is gratifying that their resulting CRs, based on 
independent data, are rather similar to ours (e.g., compare their 
Figure~\ref{Fig:eta-P_L0} with our 
Figure~\ref{Fig:H-Omega_M_L0_G+-0.05}).

\end{document}